\documentclass[letterpaper, aps, prd, twocolumn, superscriptaddress, showpacs, nofootinbib]{revtex4}
\pdfoutput=1

\usepackage{graphicx}
\usepackage{bm}
\usepackage{dcolumn}
\usepackage{color}
\usepackage[sort&compress]{natbib}
\usepackage[latin9]{inputenc}
\usepackage{url}
\usepackage{subfigure}
\usepackage{float}
\usepackage{longtable}
\usepackage{amssymb}
\usepackage{amsmath}
\usepackage{amsfonts}
\begin{document}

\title{Astrophysics with core-collapse supernova gravitational wave signals in the next generation of gravitational wave detectors}

\newcommand*{\vinnyuni}{University of Oregon, Eugene, Oregon, 97403, USA.}
\affiliation{\vinnyuni}
\newcommand*{\ozgrav}{OzGrav, Swinburne University of Technology, Hawthorn, Melbourne, VIC 3122, Australia.}
\affiliation{\ozgrav}
\newcommand*{\glasgow}{University of Glasgow, Physics and Astronomy, Kelvin Building, Glasgow, Lanarkshire, G128QQ, UK.}
\affiliation{\glasgow}

\author{Vincent Roma} \affiliation{\vinnyuni}
\author{Jade Powell}  \affiliation{\ozgrav}
\author{Ik Siong Heng} \affiliation{\glasgow}	
\author{Raymond Frey} \affiliation{\vinnyuni}

\date{\today}

%%%%%%%%%%%%%%%%%%%%%%%%%%%%%%%%%%%%%%%%%%%%%%%%%%%%%%%%%%%%%%%%%%%%
%%%%%%%%%%%%%%%%%%%%%%%%%%%%%%%%%%%%%%%%%%%%%%%%%%%%%%%%%%%%%%%%%%%%
\begin{abstract} 
The next generation of gravitational wave detectors will improve the detection prospects for gravitational waves from core-collapse supernovae. The complex astrophysics involved in core-collapse supernovae pose a significant challenge to modeling such phenomena. The Supernova Model Evidence Extractor (SMEE) attempts to capture the main features of gravitational wave signals from core-collapse supernovae by using numerical relativity waveforms to create approximate models. These models can then be used to perform Bayesian model selection to determine if the targeted astrophysical feature is present in the gravitational wave signal. In this paper, we extend SMEE's model selection capabilities to include features in the gravitational wave signal that are associated with g-modes and the standing accretion shock instability. For the first time, we test SMEE's performance using simulated data for planned future detectors, such as the Einstein Telescope, Cosmic Explorer, and LIGO Voyager. Further to this, we show how the performance of SMEE is improved by creating models from the spectrograms of supernova waveforms instead of their time-series waveforms that contain stochastic features. In third generation detector configurations, we find that about 50\% of neutrino-driven simulations were detectable at 100 kpc, and 10\% at 275 kpc. The explosion mechanism was correctly determined for all detected signals. 
\end{abstract}

\maketitle

%%%%%%%%%%%%%%%%%%%%%%%%%%%%%%%%%%%%%%%%%%%%%%%%%%%%%%%%%%%%%%%%%%%%
%%%%%%%%%%%%%%%%%%%%%%%%%%%%%%%%%%%%%%%%%%%%%%%%%%%%%%%%%%%%%%%%%%%%
\section{Introduction}
\label{sec:intro}

Current ground based gravitational wave detectors, such as Advanced LIGO (aLIGO)~\cite{aLIGO} and Advanced Virgo (AdVirgo)~\cite{AdVirgo}, are sensitive to gravitational waves emitted by core-collapse supernovae (CCSNe)~\cite{2016arXiv160501785A, 2016PhRvD..93d2002G}. It is expected that the advanced detectors could detect these sources out to distances of a few kpc~\cite{2016PhRvD..93d2002G, 2017PhRvD..96l3013P}. However, the rates for CCSNe are low at these distances~\cite{1991ARA&A..29..363V, 1993A&A...273..383C, Alexeyev2002}, indicating that a third generation of gravitational wave detectors may be needed to make the first CCSN gravitational-wave detections. 

Current proposed third generation detectors include the Einstein Telescope (ET)~\cite{0264-9381-27-19-194002}, that will consist of three underground 10\,km interferometers in a triangular geometry, KAGRA~\cite{2013PhRvD..88d3007A}, a 3\,km interferometer currently under construction in the Kamioka mine in Japan, and Cosmic Explorer~\cite{2017CQGra..34d4001A}, a 40\,km detector. Other possible future detectors may include proposed upgrades to aLIGO, such as LIGO Voyager~\cite{2015PhRvD..91h2001D}. Current astrophysics studies of third generation detectors have focused on compact binary sources~\cite{2018PhRvD..97l3014C, 2018PhRvD..97f4031Z, 2017PhRvD..95f4052V}, or a stochastic gravitational wave background~\cite{2017PhRvD..95f3015C}. However, understanding the CCSN science that can be performed with third generation detectors is important for informing the science case and design of the instruments.  

Gravitational wave predictions produced in 2D CCSN simulations have been extensively used to investigate the ability to measure astrophysical parameters of CCSN gravitational wave signals. This includes how rapidly the star is rotating~\cite{2014PhRvD..90d4001A}, the equation of state (EOS)~\cite{2017PhRvD..95f3019R}, and the explosion mechanism of the star~\cite{logue:12, powell:16b}. However, CCSN simulations have advanced rapidly in recent years. A number of gravitational wave predictions produced in 3D CCSN simulations are now available~\cite{mueller:e12, scheidegger:10b, kuroda:16, andresen:16, 2017arXiv170107325Y, 2018arXiv181007638A}. Their waveform predictions show significant differences to the 2D case, and some common features have been identified between different simulation groups. Rapidly rotating waveforms have identified a spike in the time series that occurs at core bounce~\cite{scheidegger:10b}. Other simulations show low frequency gravitational wave emission due to the standing accretion shock instability (SASI)~\cite{2015PhRvD..92h4040Y, 2013ApJ...766...43M}, and higher frequency gravitational wave emission due to g-modes at the surface of the proto-neutron star (PNS)~\cite{2013ApJ...766...43M}. The identification of these features has enabled studies towards asteroseismology with gravitational wave observations~\cite{2018MNRAS.474.5272T, 2018arXiv180611366T}. 

Previous studies of CCSN signals have focused on ground based detectors. Mock data studies have investigated how well CCSN signals can be detected in aLIGO, AdVirgo and Kagra~\cite{2015PhRvD..92l2001H, 2016PhRvD..93d2002G}, and how the sensitivity can be improved with a Bayesian classification of events~\cite{2018arXiv180207255G}. Other studies have applied principal component analysis (PCA) to determine the explosion mechanism of CCSN signals~\cite{heng:09, 2014InvPr..30k4008E, logue:12, powell:16b, 2017PhRvD..96l3013P}. The Supernova Model Evidence Extractor (SMEE) is a Bayesian model selection pipeline that applies PCA to time-series CCSN gravitational wave predictions to produce approximate models to represent gravitational wave emission expected from different astrophysical features. Previous SMEE studies have focused on a one detector proof-of-principal case~\cite{logue:12}, an aLIGO and AdVirgo realistic noise study~\cite{powell:16b}, and using the time series of new 3D CCSN simulations~\cite{2017PhRvD..96l3013P}. 

In this paper, we investigate the performance of SMEE in the next generation of gravitational wave ground based detectors. Further to this, we update SMEE to create signal models from spectrograms of predicted CCSN signals, as the time series contains stochastic features that are difficult to predict, limiting SMEE's model selection capabilities. As features like g-modes have specific signatures in a spectrogram, this allows us to extend SMEE's available models to include models that represent the gravitational wave emission expected from SASI and g-modes.   

This paper is structured as follows: In Section~\ref{sec:waveforms}, we describe the gravitational wave signals used in this analysis. In Section~\ref{sec:smee}, we give a brief overview of SMEE with a detailed description of changes to the algorithm since the last study. In Section~\ref{sec:detectors}, we describe the gravitational wave detectors considered and their sensitivity. In Section~\ref{sec:method}, we outline the analysis. The results are given in Section~\ref{sec:results}, and a conclusion and discussion is given in Section~\ref{sec:conclusion}.   

%%%%%%%%%%%%%%%%%%%%%%%%%%%%%%%%%%%%%%%%%%%%%%%%%%%%%%%%%%%%%%%%%%%%
%%%%%%%%%%%%%%%%%%%%%%%%%%%%%%%%%%%%%%%%%%%%%%%%%%%%%%%%%%%%%%%%%%%%
\section{Gravitational wave signals from core-collapse supernovae}
\label{sec:waveforms}

\begin{figure}[!ht]
\centering
\includegraphics[width=\columnwidth]{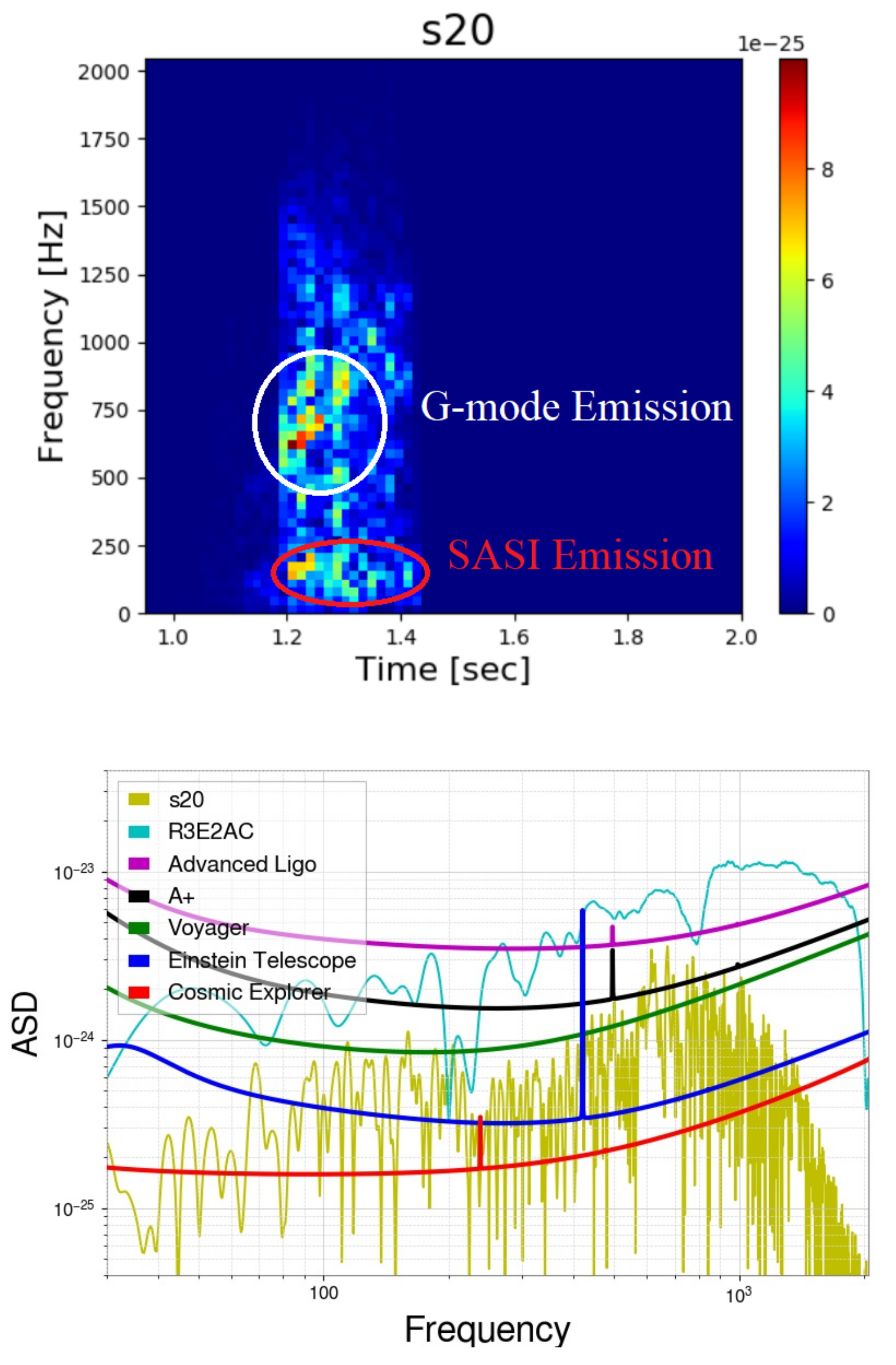}
\caption{Top: Spectrogam of a neutrino model gravitational wave signal for a $20\,M_{\odot}$ progenitor star simulated by Andresen et al.~\cite{andresen:16}. The g-mode and SASI waveform features have been circled in white and red respectively. Core bounce occurs at t = 1.0\,s. Bottom: ASD of the same neutrino model waveform and magnetorotational waveform (R3E2AC) from Scheidegger et al.~\cite{scheidegger:10b} plotted with projected future detector noise curves. Both waveforms represent a source distance of 50\,kpc and use a sky-averaged antenna pattern of .44\,.}
\label{fig:example}
\end{figure}

\begin{table*}[htb!]
\begin{center}
\begin{tabular}{|c|c|c|c|c|c|c|c|} %{|m{3cm} m{6cm}|}
 \hline
Model Name & Mass [$M_{\odot}$] & Duration [s] & Peak Freq [Hz] & Max Amplitude at 10kpc & g-Modes  & SASI \\ 
 \hline
s11 & 11 ZAMS & .345 & 550 & 8e-23 & Yes & No \\ 
 \hline
s20 & 20 ZAMS & .421 & 100 & 2e-22 & Yes & Yes \\
 \hline
s20s & 20 ZAMS & .529 & 100 & 2.5e-22 & Yes & Yes \\
 \hline
s27 & 27 ZAMS & .575 & 100 & 1.5e-22 & Yes & Yes \\
 \hline
sfhx & 15 ZAMS & .350 & 671 & 6e-22 & Yes & Yes \\
 \hline
tm1 & 15 ZAMS & .350 & 635 & 4e-22 & Yes & No \\
 \hline
L15 & 15 ZAMS & 1.4 & 160 & 1e-22 & No & Yes \\
 \hline
W15 & 15 ZAMS & 1.3 & 192 & 1.1e-22 & No & Yes \\
 \hline
N20 & 20 ZAMS & 1.5 & 96 & 7e-23 & No & Yes \\
 \hline
he3.5 & 3.5 He star & .627 & 850 & 2e-22 & Yes & No \\
 \hline
s18 & 18 ZAMS & .625 & 850 & 4e-22 & Yes & No \\
 \hline
C15 & 15 ZAMS & .436 & 1000 & 9e-22 & Yes & Yes \\
 \hline
 \end{tabular}
\caption{An overview of the neutrino-driven mechanism waveforms used in this study. The waveforms differ considerably in their amplitude, frequency, and duration. For this study, we only consider g-modes which occur above 300\,Hz. }
\label{tab:neutrino_waves}
\end{center}
\end{table*}

In this section, we summarize a few common features of CCSN gravitational wave emission and give a brief overview of the CCSN waveforms used in this paper.  %In future sections, we refer to the Scheidegger waveforms as the magnetorotational catalog and we combine all the waveforms that explode by the neutrino mechanism into a new waveform catalog that we refer to as the neutrino catalog.

Magnetorotational simulations are usually dominated by prompt broadband emission right at core bounce. This is because the core of a rapidly rotating star will be deformed due to its angular momentum, causing derivatives of the quadrupole moment to change significantly as the core collapses~\cite{scheidegger:10b}. This core bounce spike is the predominant feature in rapidly rotating simulations.

Table~\ref{tab:neutrino_waves} gives an overview of the gravitational wave emission features found in neutrino-driven waveforms. Neutrino-driven waveforms are typically simulated without rotation and therefore do not posses a spike at core bounce. Instead their emission typically grows over a few hundred milliseconds as features such as SASI and g-modes increase in strength. After core bounce, a shock front will propogate through the star disassociating infalling matter~\cite{muller2016core}. Low $(l, m)$ oscillations of this shock front are known as SASI. Typically the amplitudes of these spherical harmonics will grow exponentially during the linear phase and then saturate in the non-linear phase~\cite{hanke2013sasi}. This results in low frequency emission as seen in Figure~\ref{fig:example}. Oscillations within the PNS are also expected to emit gravitational waves and these are known as g-modes, or gravity modes. These g-modes have buoyancy as a restoring force and can exist within the inner region of a neutron star or at the surface~\cite{andresen:16}. These modes can be instigated or influenced by downflows of matter impinging upon the surface of the PNS~\cite{kuroda:16}. Gravitational wave emission from g-modes typically grows in frequency over time as seen in Figure~\ref{fig:example}. SASI and g-modes are two of the most prominent features to appear in the gravitational wave emission predicted in CCSN simulations. Confirming their existence in nature would benefit our understanding of the CCSN explosion process and serve as confirmation of the gravitational wave predictions produced in recent simulations. The work from different simulation studies used in this paper is described below.

\subsection{Scheidegger}

Scheidegger et al.~\cite{scheidegger:10b} produce 25 gravitational waveforms from 3D magnetohydrodynamic (MHD) core-collapse simulations of a $15\,M_{\odot}$ zero age main sequence (ZAMS) progenitor star. The spectra of a typical Scheidegger waveform can be seen in Figure~\ref{fig:example}. They use the Lattimer-Swesty and Shen equation of states (EOS), and a variety of rotation values from non-rotating to rapidly-rotating. We use only the 15 most rapidly rotating waveforms. The rotation leads to a large spike at core-bounce in the plus polarization only. The simulations are short duration as they were stopped up to 130\,ms after the core bounce time. 

\subsection{M\"uller}

M\"uller et al.~\cite{mueller:e12} carry out neutrino-driven CCSN simulations of non-rotating stars. They are 3D simulations that produce two gravitational wave polarizations. There are two models with a $15\,M_{\odot}$ ZAMS progenitor star, referred to as models L15 and W15, and one model with a $20\,M_{\odot}$ progenitor star, referred to as model N20. The waveforms have emission due to both SASI and g-modes. The lower frequency gravitational wave emission is extended as they artificially prescribed the contraction of the neutron star. The gravitational wave signals extend to 1.3\,s after core bounce, with the strongest gravitational wave emission in the first 0.7\,s after the core bounce time. The M\"uller et al. waveforms do have g-mode emission, however it is relatively slow to develop and rarely approaches 300\,Hz~\cite{mueller:e12}. For the purposes of this analysis, we focus our g-mode analysis on the higher frequency g-modes that are present in the more recent simulations~\cite{andresen:16, kuroda:16}.  

\subsection{Andresen}

Andresen et al.~\cite{andresen:16} also carry out 3D neutrino-driven CCSN simulations of non-rotating stars. They produce four gravitational wave signals, two with a $20\,M_{\odot}$ progenitor star, referred to as models s20 and s20s, one with an $11\,M_{\odot}$ progenitor, referred to as model s11, and one with a $27\,M_{\odot}$ progenitor, referred to as model s27. Model s20 does not successfully revive the shock, and model s20s does experience shock revival. The $20\,M_{\odot}$ model is shown in Figure~\ref{fig:example}. There is strong low frequency gravitational wave emission due to the SASI in all the Andresen models except model s11. The gravitational wave emission above 1000\,Hz is not reliable due to an aliasing problem due to a low sample rate. The duration of the signals varies from 350\,ms to 600\,ms after core bounce. 

\subsection{Kuroda}

Kuroda et al.~\cite{kuroda:16} carry out 3D simulations of a $15\,M_{\odot}$ ZAMS progenitor star using three different EOS. We use the two models tm1 and sfhx that correspond to two different EOS. The simulations were stopped 340\,ms after core bounce. Both models show gravitational wave emission that originates with g-mode oscillations of the PNS surface. Model sfhx also shows gravitational wave emission at lower frequencies, as the model experiences sloshing and spiral motions of the SASI before neutrino-driven convection dominates.    

\subsection{Yakunin}

Yakunin et al.~\cite{2017arXiv170107325Y} carry out one general relativistic, multi-physics, 3D simulation of a $15\,M_{\odot}$ ZAMS progenitor star with state of the art weak interactions. The simulation is stopped 450\,ms after core bounce. The strong gravitational wave emission starts at $\sim120$\,ms after core bounce when the SASI develops. This model has larger gravitational wave amplitudes, up to $\sim20$\,cm, than other recent neutrino-driven simulations and the emission peaks at a higher frequency of 1000\,Hz due to g-mode oscillations of the PNS surface. 

\subsection{Powell}

Powell et al.~\cite{2018arXiv181205738P} carry out two neutrino-driven simulations in 3D down to the innermost 10\,km to include the PNS convection zone in spherical symmetry. The first simulation is the explosion of an ultra-stripped star in a binary system simulated from a star with an initial helium mass of $3.5\,M_{\odot}$, which we refer to as model he3.5. The ultra-stripped simulation ends at 0.7\,s after core bounce. The second is a single star with a ZAMS mass of $18\,M_{\odot}$, which was simulated up to 0.9\,s after core bounce and we refer to as model s18. Both models have peak gravitational wave emission between 800\,Hz and 1000\,Hz due to g-mode oscillations of the PNS surface. The peak amplitude of model he3.5 is $\sim6$\,cm and the peak amplitude of the s18 model is $\sim10$\,cm. We only use the first 0.62\,s for each of the Powell models as we began our study before their simulations were completed. However, both models have their peak gravitational wave emission at earlier times than 0.6\,s, so not much gravitational wave emission is lost by using shorter versions of the waveforms.

%%%%%%%%%%%%%%%%%%%%%%%%%%%%%%%%%%%%%%%%%%%%%%%%%%%%%%%%%%%%%%%%%%%%
%%%%%%%%%%%%%%%%%%%%%%%%%%%%%%%%%%%%%%%%%%%%%%%%%%%%%%%%%%%%%%%%%%%%
\section{The supernova model evidence extractor}
\label{sec:smee}

This section will summarize the analysis method implemented in SMEE, with an emphasis on changes from previous publications. % In previous versions, PCs were created in the time domain, the signal model was constructed out of PCs in the FFT domain, and a Gaussian likelihood was used for the signal and noise model likelihoods.

\subsection{Principal Component Analysis (PCA)}

PCA is performed on a catalog of waveforms to create a set of orthogonal basis vectors called Principal Components (PCs). The PCs are ranked by the amount of variance described in each PC, with the most important waveform features contained in the first few PCs~\cite{logue:12, powell:16b, 2017PhRvD..96l3013P, heng:09}.
%The PCs contain the most important features of the waveforms in a catalog~\cite{logue:12, powell:16b, 2017PhRvD..96l3013P, heng:09}. 
Before creating the PCs, the signals used in this study are converted to three second long zero-padded amplitude spectral density (ASD) spectrograms. Each ASD within the spectrogram contains .03125 seconds of data and we use a 50\% overlap. In previous versions of SMEE, PCs were created from Fourier-transformed time series waveforms. However, time series CCSN gravitational wave signals have stochastic components. These unpredictable features are mitigated in a spectrogram making the results more robust. Each successive ASD in a spectrogram is a vector, containing one power value for each frequency bin. We attach data from the second ASD to the back of the first, and so on, to create a single vector out of each spectrogram. We then create a matrix $\textbf{D}$, where each column contains a waveform. By performing Singular Value Decomposition (SVD) on the matrix $\textbf{D}$, the data can be factored such that,
\begin{equation}
\textbf{D} = \textbf{U}\,\boldsymbol\Sigma\,\textbf{V}^{T}\,,
\end{equation}
The columns of $\textbf{U}$ and $\textbf{V}$ contain the eigenvectors of $\textbf{D}\textbf{D}^{T}$ and $\textbf{D}^{T} \textbf{D}$, respectively~\cite{heng:09}. $\boldsymbol\Sigma$ is a diagonal matrix with elements corresponding to the square roots of the eigenvalues. The PCs are the orthonormal eigenvectors in $\textbf{U}$. The PCs are ranked by their eigenvalues in $\boldsymbol\Sigma$, meaning that the first few PCs contain the most important features of a catalog. A linear combination of PCs is used to construct our signal model, 
\begin{equation}
h_i \approx \left| \sum_{j=1}^{k} U_j \beta_j \right| \,,
\label{equ:linear_sum}
\end{equation}
where $h_i$ is our reconstructed waveform, $U_j$ is the $j$th PC, $\beta_j$ is the corresponding PC coefficient, and $k$ is the number of PCs being used. This reconstruction can then be used as a model for Bayesian model selection.

% \subsection{Supernova PC models}

% \begin{figure*}[p]
% \centering
% \includegraphics[width=\textwidth,height=\textheight]{Figures/PC_plot.PNG}
% \caption{The first three PCs for each catalog. From top to bottom, the catalogs are: Neutrino mechanism, Magnetorotational mechanism, g-modes, No g-modes, SASI, No SASI.}
% \label{fig:pcs}
% \end{figure*}

% In this study, we aim to perform model selection to determine the explosion mechanism, to determine if g-modes are present in the signal, and to determine if emission from SASI is present in the signal. This means that we need to apply PCA to several different catalogs of waveforms to produce signal models for this study. Figure \ref{fig:pcs} shows the first three PCs for all the signal models used in this study. The first is the PCs for the neutrino mechanism model. They were created from \jade{?} neutrino mechanism waveforms described in Section \ref{sec:waveforms}. The second is the magnetorotational catalog created from the 15 Scheidegger waveforms also described in section \ref{sec:waveforms}. The other PCs are for models used to determine if SASI and g-modes are present in the signal. 

% Even though these PCs contain both positive and negative values, the signal model will always be a positively-valued amplitude spectral density (ASD) spectrogram due to the absolute value around the linear sum in equation \ref{equ:linear_sum} \cite{logue:12}. The first few PCs contain the most prominent waveform features of a catalog. 

\subsection{Bayesian Model Selection}
\label{subsec:bayesian}

Bayesian model selection is used to calculate Bayes factors that allow us to distinguish between two competing models. The Bayes factor $B_{S,N}$ is given by the ratio of the evidences~\cite{veitch:15},
\begin{align}
B_{S,N} = \frac{p(d|M_{S})}{p(d|M_{N})}\,,
\end{align}
where $d$ is our data and $M_{S}$ and $M_{N}$ are the signal and noise models, respectively. The evidence is given by the integral of the likelihood multiplied by the prior across all parameter values $\theta$~\cite{logue:12, powell:16b}.
\begin{equation}
\label{equ:integral}
p(d|M) = \int_{\theta} p(\theta|M)p(d|\theta,M)d\theta
\end{equation}
This integral can be solved via nested sampling~\cite{skilling2004nested}. Specifically, SMEE uses the nested sampler within LALInference~\cite{veitch2010bayesian, veitch:15}.

The parameters for the models in SMEE are the root sum squared amplitude of the signal $h_{rss}$~\cite{2016arXiv160501785A}, the arrival time, the polarization angle, and the PC coefficients. The prior for $h_{rss}$ is uniform in volume and bounded between $1.3\times10^{-24}$ and $5.8\times10^{-19}$. The time prior is uniform across a one second interval. We make no assumptions about the polarization angle, so its prior is uniform between 0 and $2\pi$. The PC coefficients also use uniform priors with the bounds determined by the maximum and minimum values necessary to reconstruct all catalog waveforms (plus or minus 25\% respectively to account for waveform uncertainty). We assume the sky position is known for the analysis in this paper.

For our analysis, we use the logarithm of the Bayes factor. When $\log B_{S,N} > 0\,$, the signal model is preferred over the noise model. When $\log B_{S,N} < 0\,$, the noise model is preferred. Two different signal models can be compared in the same way via their $\log B_{S,N}$ values~\cite{veitch:15},
\begin{equation}
\log B_{ij} = \log p(d|M_{i}) - \log p(d|M_{j})
\end{equation}
If each signal model corresponds to a different CCSN explosion mechanism, then we can select the emission model that best reconstructs the observed CCSN waveform.

\subsection{Signal and Noise Likelihoods}

Phase information in CCSN waveforms is difficult to model accurately due to the inherent stochastic nature of the explosions. Because of this, our spectrograms are entirely real-valued and do not contain any phase information. The power in each frequency bin is found by summing the squares of the real part with that of the imaginary part. If the real and imaginary parts are each Gaussian variables, this leads to a noncentral chi-squared distribution ($l = 2$) for our likelihood~\cite{logue:12},
\begin{multline}
\log \mathcal{L}_{S} = N\log\left(\frac{1}{2}\right) + \\ \sum_{n=1}^{N} \left( \frac{-(d_{n}^{2} + h_{n}^{2})}{S(f)} +  \log\left(I_0\left[\frac{2 d_{n} h_{n}}{S(f)}\right]\right) \right)
\end{multline}
where $d$ is a one-sided ASD of the detector data, $h$ is the signal model constructed with the PCs, $S(f)$ is the one-sided noise power spectral density (PSD), and $N$ is the total number of data points (frequency bins) with corresponding index $n$. The analysis is performed coherently, meaning data points from all detectors are used in the same calculation. The likelihood function gives the probability of the data consisting of the signal model plus Gaussian noise. The noise only likelihood function is identical, except $h = 0$~\cite{logue:12},
\begin{equation}
\log \mathcal{L}_{N} = N\log(1/2) - \sum_{n=1}^{N} \frac{d_{n}^{2}}{S(f)} .
\end{equation}
This tests the data's consistency with Gaussian noise. Because we are concerned with likelihood ratios, the constant terms are dropped in both signal and noise models~\cite{logue:12, powell:16b}.

\subsection{Signal Model Catalogs}
\label{subsec:pcs}

\begin{figure*}[p]
\includegraphics[height=\textheight]{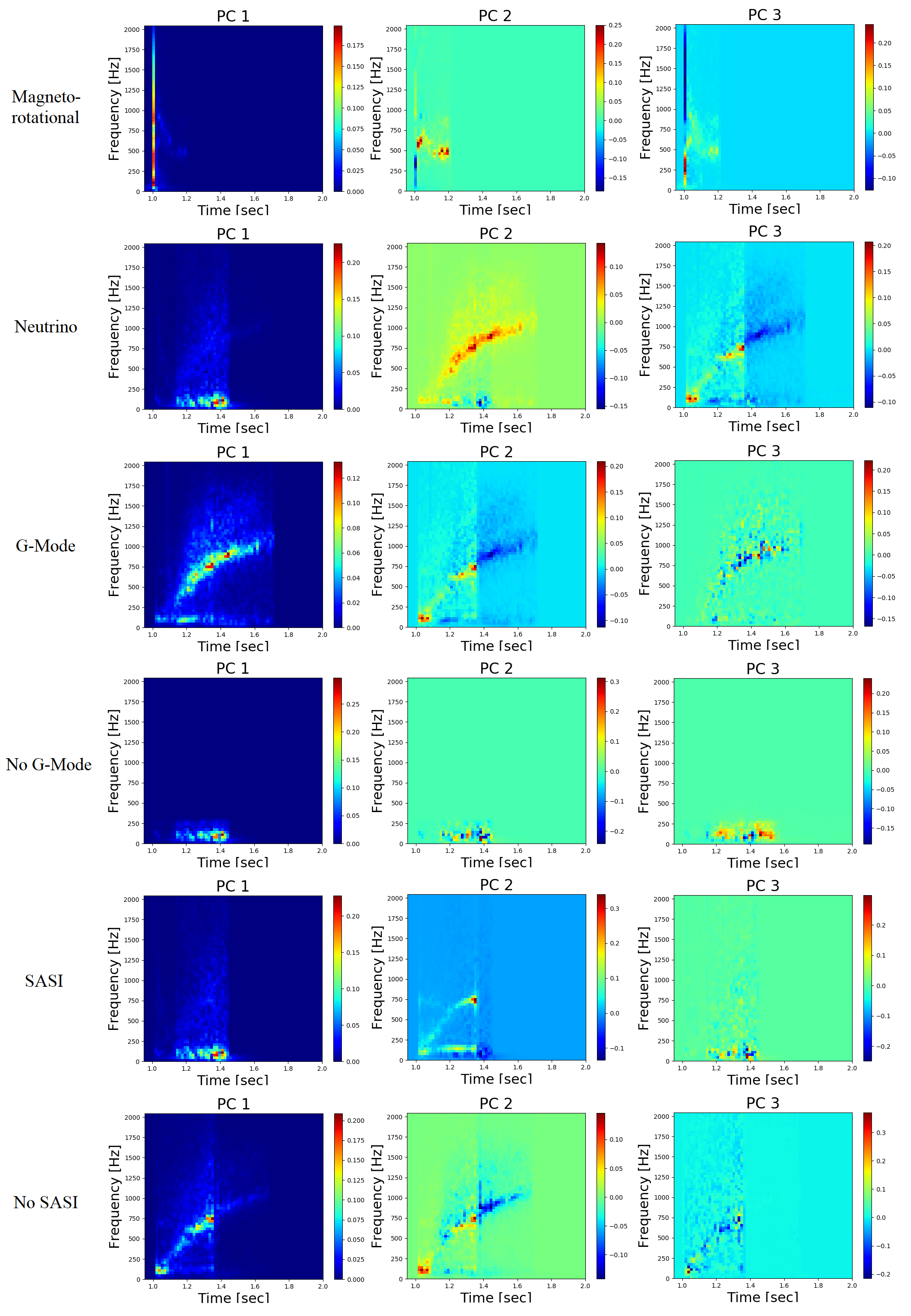}
\caption{The first three PCs for each catalog. From top to bottom, the catalogs are: neutrino mechanism, magnetorotational mechanism, g-modes, no g-modes, SASI, and no SASI. The principal components represent the most important waveform features for each catalog.}
\label{fig:pcs}
\end{figure*}

The aim of this study is to produce three different classification statements about detections of gravitational waves from CCSN in future detectors. The first is the most likely explosion mechanism (neutrino vs magnetorotational), and the others are the presence of specific features in the waveform that are associated with g-modes or SASI. Each statement requires two or more signal models to be compared to each other in order to produce a Bayes factor. For explosion mechanism classification, one signal catalog contains 10 waveforms pertaining to the neutrino model, and the other catalog contains 13 waveforms pertaining to the magnetorotational model. Both polarizations are used for all simulated waveforms. All of the magnetorotational waveforms come from the Scheidegger group while the neutrino models come from the Andresen, Mueller, Kuroda, Powell, and Yakunin groups described in Section~\ref{sec:waveforms}. PCs are produced for each signal catalog. The first few PCs are linearly combined to produce models that are used to calculate Bayes factors that signify the more likely mechanism. 

Figure~\ref{fig:pcs} shows the first three PCs for all the signal models used in this study. Even though the PCs contain both positive and negative values, the signal model will always be a positively-valued amplitude spectral density (ASD) spectrogram due to the absolute value around the linear sum in Equation~\ref{equ:linear_sum}~\cite{logue:12}. The first few PCs contain the most prominent waveform features of a catalog. There are 2 waveforms left out of each catalog to be used later for realistic performance testing. Model s20s, which exhibits features associated with SASI, was omitted from the neutrino catalog, along with model sfhx, which exhibits strong g-mode emission. For the magnetorotational catalog, one waveform with a low $h_{rss}$ (R3E1AC) and one with a high $h_{rss}$ (R4E1FC\_L) were omitted from the construction of the PCs.

Catalogs were set up similarly to make statements about g-modes and SASI, with one catalog's waveforms possessing the feature and the other catalog's waveforms not possessing the feature. The PCs for these models are also shown in Figure~\ref{fig:pcs}. It should be noted that all of the waveforms used in the g-mode and SASI catalogs were non-rotating neutrino model waveforms~\cite{mueller:e12, kuroda:16, andresen:16, 2017arXiv170107325Y}. These features are typically not present in the currently available rotational models due to reduced convection and because their simulations were stopped shortly after the core bounce time. The g-mode catalog contains 6 waveforms from models s20, s20s, s27, he3.5, s18, and tm1. Model sfhx was left out as a non-catalog waveform for testing. The no g-mode catalog contains 5 waveforms from models s20, s20s, s27, N20, and C15. The model L15 was left out as a non-catalog waveform for testing. The Andresen and Yakunin waveforms for this specific catalog were lowpass filtered at 255\,Hz to remove their higher frequency g-mode signals while leaving behind their SASI emission. The SASI catalog contains 6 waveforms from models s20, s27, sfhx, L15, N20, and C15. The model s20s was the non-catalog waveform used for testing. The no SASI catalog contains 3 waveforms from the models s11, he3.5, and tm1. The s18 model was left out as a non-catalog waveform for testing.

\begin{figure*}[ht]
\centering
\includegraphics[width=\columnwidth]{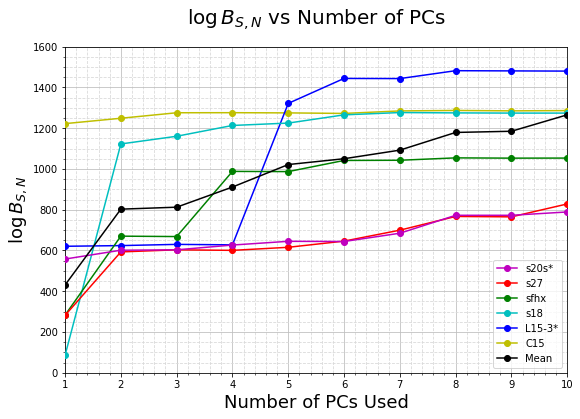}
\includegraphics[width=\columnwidth]{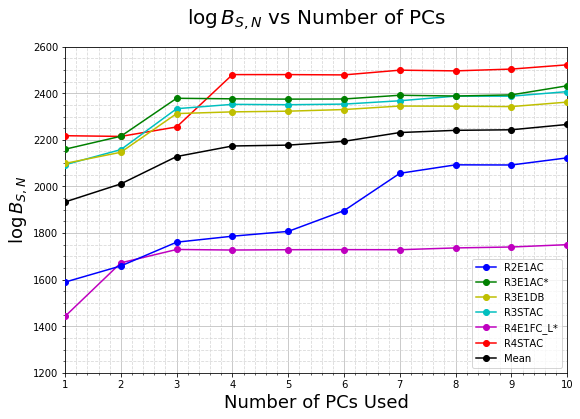}
\caption{Log Bayes factors for mechanism classification with an increasing number of PCs in simulated aLIGO Gaussian noise. Waveforms with a * were not included when making the PCs. The left figure shows neutrino catalog waveforms, and the right figure shows magnetorotational catalog waveforms. The log Bayes factors stop increasing  when an ideal number of PCs is reached. We chose 5 PCs for the neutrino model and 4 for the magnetorotational as there was limited $\log B_{S,N}$ improvement beyond those points.}
\label{fig:pickpcs}
\end{figure*}

Different waveform catalogs can have different levels of complexity and variation, resulting in different numbers of PCs needed to accurately reconstruct signals~\cite{logue:12, powell:16b, 2017PhRvD..96l3013P}. Additionally, Bayesian model selection tends to prefer simpler models that require fewer PCs, which can affect results in some cases such as when the SNR of the gravitational wave signal is low. We determine the optimal number of PCs by injecting and reconstructing multiple waveforms from each catalog using an increasing number of PCs. We then consider $\log B_{S,N}$ vs number of PCs to determine the point at which most waveforms have reached their maximum $\log B_{S,N}$. The results using our mechanism classification models are shown in Figure~\ref{fig:pickpcs}. We choose 5 PCs for the neutrino model and 4 for the magnetorotational model as performance improvement is limited beyond those numbers. Previous versions of SMEE used 5 and 8 PCs for magnetorotational and neutrino models respectively~\cite{2017PhRvD..96l3013P}. The decline in the required number of PCs is likely due to the robustness of the spectrogram format and happens in spite of the fact that we are now using 3 additional neutrino waveforms since the last SMEE publication. We are also now using both polarizations in the PCA. The waveform feature catalogs produced similar results to Figure~\ref{fig:pickpcs}. We use 5 PCs for the no g-mode and no SASI catalogs and 8 PCs for the g-mode and SASI catalogs. Waveforms within the last two catalogs can contain both waveform features and therefore more complexity is needed to ensure good performance.

%%%%%%%%%%%%%%%%%%%%%%%%%%%%%%%%%%%%%%%%%%%%%%%%%%%%%%%%%%%%%%%%%%%%%%
%%%%%%%%%%%%%%%%%%%%%%%%%%%%%%%%%%%%%%%%%%%%%%%%%%%%%%%%%%%%%%%%%%%%%%
\section{Gravitational wave detectors}
\label{sec:detectors}

In this section, we give a brief overview of the future gravitational wave detectors considered in this study. 

\begin{figure}[ht]
\centering
\includegraphics[width=\columnwidth]{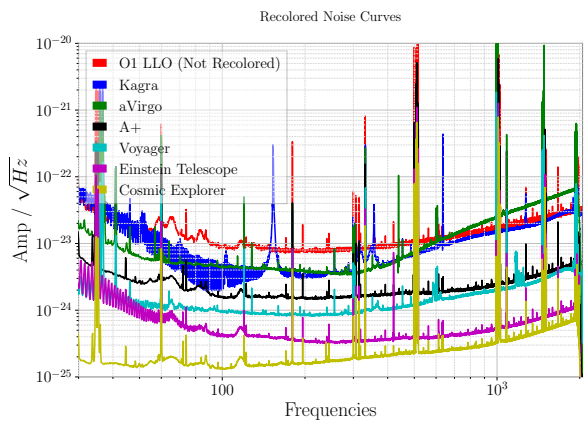}
\caption{Spectra for LIGO O1 data recolored to future detector sensitivities. Data is shown from 30 - 2048 Hz, the frequency band used in SMEE's analysis.}
\label{fig:recolored_curves}
\end{figure}

\subsection{A+}

LIGO A+ is a relatively modest set of planned upgrades to the aLIGO detectors in Livingston, LA and Hanford, WA~\cite{miller2015prospects}. 
%These upgrades are being implemented in a series of incursions, with the intent of minimizing detector downtime. 
The improvements of A+ are focused on aLIGO's dominant noise sources, quantum noise and coating thermal noise. This potentially includes the installation of a squeezed light source with a filter cavity, and new end-test-masses (and possibly input-test-masses) with improved coatings. Low-risk upgrades to the suspensions may also be made, such as modifications to reduce gas damping, improve bounce and roll mode damping, mitigate parametric instabilities, etc~\cite{whitepaper2018}.

\subsection{Advanced Virgo}

Virgo is a 3\,km interferometer operating out of Italy~\cite{AdVirgo}. It was initially constructed in 2003 and has recently undergone various upgrades to become AdVirgo. AdVirgo began commissioning in 2016 and participated in the second observing run (O2) with the aLIGO detectors. AdVirgo is performing three major upgrades in preparation for O3. It is currently upgrading its input laser to higher power, replacing its steel test mass suspension wires with ones made of fused silica, and adding a squeezed vacuum source at the output of the interferometer. These upgrades should result in AdVirgo approaching design sensitivity~\cite{AdVirgo, whitepaper2018}.

\subsection{Kagra}

Kagra is a 3\,km long underground interferometer built in the Kamioka mine in Japan~\cite{2013PhRvD..88d3007A, akutsu2018kagra}. It is similar in design to other existing detectors but is built underground to minimize seismic noise, gravity gradient noise, and other environmental fluctuations such as temperature and humidity~\cite{Somiya2012detector}. aLIGO and AdVirgo both use fused silica test masses at room temperature, while Kagra uses sapphire test masses at 20\,K to reduce thermal noise. Kagra will serve as a test case and pioneer for future detectors that also incorporate underground construction and cryogenic cooling. Kagra is expected to begin observing runs as early as 2020~\cite{2013PhRvD..88d3007A, Somiya2012detector, aso2013interferometer}.

\subsection{Voyager}

LIGO Voyager is a proposed upgrade to the aLIGO detectors based on LIGO's `Blue' design concept~\cite{voyagerupgrade}. The detectors are expected to be operational by 2030~\cite{whitepaper2018}. LIGO Voyager's main modifications may include 120 - 200 kg Silicon test masses, amorphous-silicon multi-layer coatings for reduced thermal noise, low temperature ($\sim123$ K) cryogenic operation of the test masses, silicon ribbons for the final stage of test mass suspension, 200\,W pre-stabilized laser with a $\sim2000$ nm wavelength, squeezed light injection in combination with a squeezed light filter cavity, and Newtonian Gravity noise subtraction by seismometer arrays and adaptive filtering~\cite{2015PhRvD..91h2001D, whitepaper2018}.

\subsection{Einstein Telescope}

The Einstein Telescope (ET) is a proposed underground detector expected to be built in Europe~\cite{0264-9381-27-19-194002}. The current design is an equilateral triangle with 10\,km long arms. Each corner of the triangle will be a detector composed of two interferometers, one optimized for operation below 30\,Hz and the other optimized for higher frequencies~\cite{ET:design}. The low-frequency interferometers will operate from approximately 1 to 250\,Hz, use optics cooled to 10\,K, and a beam power of about 18\,kW in each arm cavity. The high-frequency interferometers will operate from 10\,Hz to 10\,kHz, use room temperature optics, and a beam power of 3\,MW. Different predicted noise curves exist for ET's different stages of development. The final predicted noise curve, ET-D, is the curve used for the analysis in this paper~\cite{ET:design}. Construction is planned to begin around mid-2021, with data potentially being recorded as early as 2027~\cite{ET:design, whitepaper2018}.

\subsection{Cosmic Explorer}

LIGO Cosmic Explorer is a 40 km detector proposed as an upgrade to Voyager~\cite{2017CQGra..34d4001A}. It will require a new site as neither of the existing sites are large enough. The technology found in Cosmic Explorer will likely be similar to that found in Voyager, but the scale will increases in key areas such as mirror mass (320\,kg), arm power (2\,Mw), and arm length (40\,km)~\cite{whitepaper2018}. It is not yet decided how many detectors there would be or where they would be located. Money for construction is expected to be awarded around 2030, with commissioning beginning by approximately 2037. If Voyager is not built then the schedule would likely be shifted forward~\cite{whitepaper2018}. For the analysis performed in this paper we used the location of aLIGO's current detector in Livingston, LA.

\section{Method}
\label{sec:method}

The analysis presented in this paper was designed to test SMEE's effectiveness in future detector arrangements. Specifically, the ability to distinguish between explosion mechanisms (neutrino vs magnetorotational) along with the ability to make statements about the presence of predicted waveform features (SASI and g-modes) in an observed signal. Future detector data is simulated as realistically as possible by adjusting segments of aLIGO's O1 data to match the estimated sensitivity curves of future detectors in a process called ``recoloring". We can then ``inject" a waveform by inserting its time series into the detector data. We inject CCSN waveforms over the entire sky at different GPS-times and with different orientations in order to test SMEE's performance on emissions from an unknown, possibly extra-Galactic source.

\subsection{Recolored Data}

Six different 24 hour segments of data from aLIGO's first observing run (O1) were chosen to be recolored. Recolored data possesses the calibration lines and transient noise glitches not found in simulated Gaussian data. The recoloring was performed with the GSTLAL software package~\cite{messick2017analysis}. GSTLAL contains software for data manipulation and the generation of simulated data. In this case, each segment of data was whitened with a reference PSD and then recolored with a filter to the noise curves of future detectors. The noise curves for the recolored data are shown in Figure~\ref{fig:recolored_curves}. Cosmic Explorer is the most sensitive detector that we consider in this study. After recoloring, each segment of data was reassigned to start at GPS time 1128211934. One segment of data, starting at 1132759948, contained O1 data from Livingston, while the other five (epochs: 1128211934, 1135238505, 1128618400, 1129037417, 1130767217) contained data from Hanford. All of the O1 data used is available from the Gravitational Wave Open Science Center (GWOSC)~\cite{2015JPhCS.610a2021V}. 

\subsection{Detector Eras}
\label{subsec:eras}

We simulated future detector configurations over the next few decades. The first configuration consists of A+ (2), AdVirgo, and Kagra. The second is similar but upgrades each LIGO detector. It has Voyager (2), AdVirgo, and Kagra. Our third configuration is identical to the previous but adds a third Voyager detector located in India. Our fourth configuration consists of the triangular ET (3) along with 2 Voyager detectors. Our fifth and final configuration consists of Cosmic Explorer and the ET (3).

\subsection{Injection and Sky Patterns}

Previous SMEE papers have examined performance from Galactic sources, usually injecting signals from the Galactic center's sky position~\cite{powell:16b, 2017PhRvD..96l3013P}, but future detectors will likely be sensitive to CCSN sources well outside of our own Galaxy~\cite{0264-9381-27-19-194002, 2017CQGra..34d4001A}. Because of this, we inject waveforms from six positions evenly distributed over the entire sky to test performance from an arbitrary source. At each sky location, injections are performed with three different polarization angles ($0, \pi/3, 2\pi/3$) evenly distributed throughout the possible parameter space. We do all of this at five different GPS times evenly distributed throughout a 24 hour period to account for Earth's rotation and the detectors' changing antenna patterns.  We inject at 5 GPS times, with 6 different sky locations, with 3 different polarization angles. That means for each signal model, each waveform is injected 90 times at each distance. This gives us a reasonable sample to gauge performance from an unknown source. Different detector eras are simulated as accurately as possible with the detectors we expect to be operational at their current expected locations and orientations.

\subsection{Mechanism Classification}

Two waveforms were omitted from the neutrino catalog and the magnetorotational catalog during the creation of the signal models. Any real life CCSN signal we observe will, of course, not identically resemble any simulated waveform in the catalog, therefore these omitted non-catalog waveforms serve as a more realistic test of a genuine signal. For the catalog waveforms, four were selected from each catalog and injected as described in the previous paragraphs. For the neutrino catalog, s20s, tm1, C15, and s18 were selected from different simulation groups. All of the magnetorotational waveforms came from the Scheidegger group, so the four chosen, R2E1AC, R3E1DB, R3STAC, and R4STAC, were evenly distributed between the minimum and maximum signal energies of the catalog. We define efficiency at a given distance as the fraction of injected waveforms that could be correctly identified with a log Bayes factor above our confidence threshold. The remaining injections could not be confidently classified. For this analysis, our confidence threshold was $\log B_{i,j} \geq 8$~\cite{powell:16b, 2017PhRvD..96l3013P}.

\subsection{Presence of SASI/g-modes}

Two non-catalog waveforms were injected for each waveform feature. One waveform has the feature and the other does not. The waveforms were injected in the manner described above and the same confidence threshold was used ($\log B_{i,j} \geq 8$). Regardless of whether the feature was present or not, the Bayes factors will be oriented such that positive answers are the correct answers. As an example, for an injection containing a g-mode signal we would calculate $\log B_{gmode,\,no\,gmode}\,$, whereas for an injection containing no g-mode signal we would calculate $\log B_{no\,gmode,\,gmode}\,$. This allows us to present related results in one simple figure.

%For the g-mode injections, Kuroda2016\_sfhx (g-mode) and M\"uller2016\_L15-3 (no g-mode) were used. For the SASI injections, Andresen2016\_s20s (SASI) and Powell2018\_s18 (no SASI) were used. These waveforms were injected in the manner described above and the same confidence threshold was used ($\log B_{i,j} \geq 8$). \jade{Techincally this is saying the non-catalog }

%%%%%%%%%%%%%%%%%%%%%%%%%%%%%%%%%%%%%%%%%%%%%%%%%%%%%%%%%%%%%%%%%%%%%%%
%%%%%%%%%%%%%%%%%%%%%%%%%%%%%%%%%%%%%%%%%%%%%%%%%%%%%%%%%%%%%%%%%%%%%%%
\section{Results}
\label{sec:results}

\subsection{Example Case Study}
\label{subsec:casestudy}

\begin{figure*}[ht!]
\centering
\includegraphics[width=15cm]{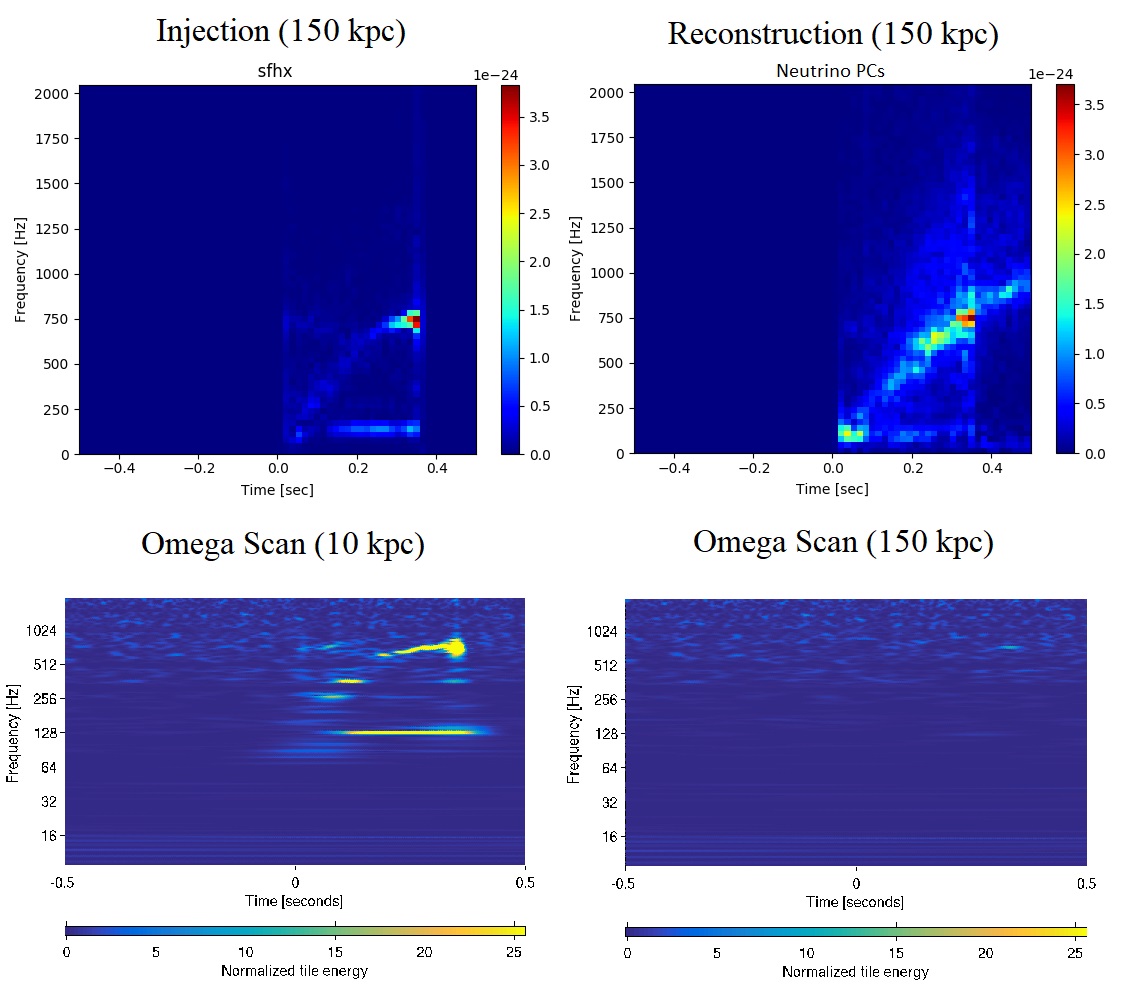}
\caption{Top row shows injected waveform and waveform reconstruction produced with the neutrino model PCs at 150\,kpc in three ET and two Voyager detectors. Bottom row shows Omega Scan spectrograms of the data from the detector in the configuration with the highest SNR. Bottom left plot shows the signal is clearly visible by eye for a 10\,kpc injection (SNR = 120), while the bottom right plot shows almost nothing visible at 150\,kpc (SNR = 8). This waveform was confidently classified at 150\,kpc as corresponding to the neutrino model with g-mode and SASI emission, even though at that distance those features are clearly not visible by eye in the noisy Omega scan.}
\label{fig:recon}
\end{figure*}

This section will summarize an example case study in which we inject model sfhx into a simulated configuration with three ET detectors and two Voyager detectors. We inject at a distance of 150\,kpc using an arbitrary sky position. The resulting network SNR was 11.4\,. The injected waveform, an example reconstruction, and example Omega Scans of one detector's data are shown in Figure~\ref{fig:recon}. Omega scans are commonly used spectrograms to view detector data within the LIGO collaboration~\cite{omegareview}. We performed all three classification statements on the signal. With the neutrino model PCs we obtained $\log B_{S,N} = 55.5\,$. With the magnetorotational PCs we obtained $\log B_{S,N} = 1.5\,$. This gives us a final mechanism classification result of $\log B_{neu,mag} = 54\,$. This result is well above our confidence threshold and tells us that our signal matches the neutrino model much more strongly than the magnetorotational. Because it appears to be a slowly rotating neutrino model waveform, we can perform the two waveform feature classification statements.

For g-mode and SASI classification, our g-mode PCs gave us a result of $\log B_{S,N} = 62.7\,$, while our no g-mode PCs gave $\log B_{S,N} = 1.3\,$. Subtracting the two gives us our final g-mode result $\log B_{gmode,\,no\,gmode} = 61.4\,$. This is also well above the confidence threshold and tells us strongly that high frequency g-modes were present in the signal. Our SASI PCs gave us $\log B_{S,N} = 97.7\,$ while our no SASI PCs gave $\log B_{S,N} = 61.8\,$. This gives us our result for SASI classification, $\log B_{SASI,\,no\,SASI} = 35.9\,$. This is also above the confidence threshold and tells us that low frequency SASI signals were present in the data. For this simulated CCSN signal, all three classification statements were made with a high confidence level even though the network SNR of the signal is relatively low and the signal features are clearly not visible in the noisy spectrogram shown in Figure~\ref{fig:recon}. % When the same model is injected with an SNR of 23, at half that distance considered in this example, almost the entire waveform is still not visible in an Omega Scan of the detector data. 

\subsection{Minimum SNR}
\label{subsec:minsnr}

\begin{figure*}[ht]
\centering
\includegraphics[width=\columnwidth]{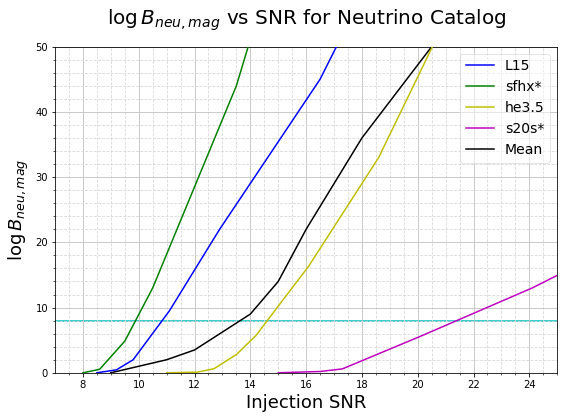}
\includegraphics[width=\columnwidth]{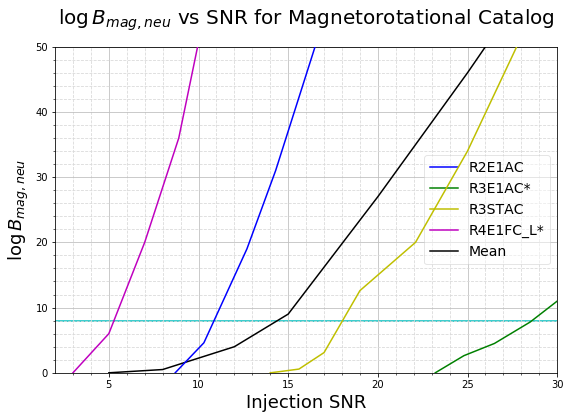}
\includegraphics[width=\columnwidth]{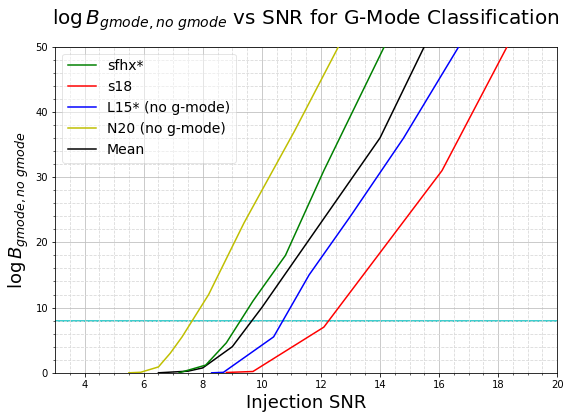}
\includegraphics[width=\columnwidth]{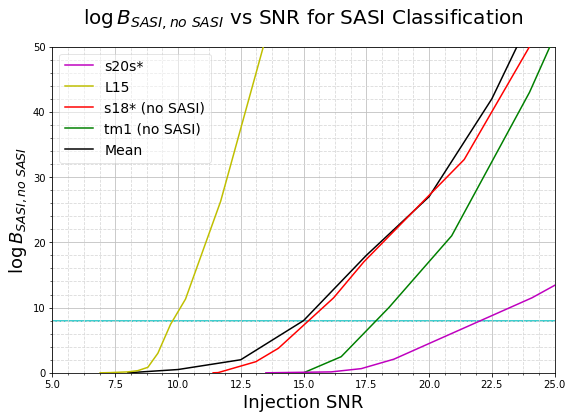}
\caption{Minimum detectable SNR for each classification statement. All injections performed in a simulated A+ configuration that included AdVirgo and Kagra. The top two plots both pertain to mechanism classification, and the bottom two are for g-modes and SASI. All results are organized such that positive Bayes values correspond to correct classifications regardless of whether the feature is present or not.}
\label{fig:min_snr}
\end{figure*}

% The minimum matched filter SNR required to classify an observed waveform depends on the statement being made. 
The top two panels in Figure~\ref{fig:min_snr} show the mechanism classification performance for neutrino and magnetorotational waveforms as a function of injected network SNR. All SNR plots were created while running SMEE on a simulated A+ configuration as described in Section~\ref{subsec:eras}. On average, neutrino model waveforms needed an SNR of about 14 or greater to be confidently classified. Some waveforms, such as s20s, required an SNR in the low twenties. This is likely due to the fact that this waveform's peak frequency of emission is around 100\,Hz and the short PSDs in our spectrograms are hindered by the 60\,Hz mains power peak present in aLIGO's data. This can lower our sensitivity to signal energy in the vicinity of 60\,Hz. Magnetorotational waveforms performed similarly and also had an average minimum SNR of about 14, with one plotted waveform requiring an SNR above 28. Despite the similar minimum SNRs, magnetorotational waveforms are generally easier to resolve at a given distance due to their larger signal energies. 
%Waveforms are typically not clearly visible by eye in detector data near their minimum classifiable SNR. This can be seen in figure \ref{fig:recon}. - JADE - I don't think we need to say this again when it's in the case study. The particular waveform from that figure was classified with an SNR of 11.4 at 150 kpc, but when injected with an SNR of 23 at half that distance almost the entire waveform was still not visible in detector data. This will vary between waveforms depending on their frequency content. --- JADE --- I don't think we need to say this again when it's in the case study. 

The bottom two panels in Figure~\ref{fig:min_snr} show SMEE's performance as a function of SNR for SASI and g-mode classification. The average minimum SNR required to make statements about the presence of g-modes was just below 9, while the average minimum for SASI statements was 15. Similarly to mechanism classification, s20s required the largest SNR of about 22 in order to confidently say that SASI was present. This is, again, likely due to our hindered sensitivity around 60\,Hz. It was generally similarly difficult for SMEE to determine that a feature was present than to rule it out, with both cases sometimes performing better.

\subsection{Mechanism Classification}
\label{subsec:mechanism_range}

\begin{figure*}[p]
\centering
\includegraphics[width=\columnwidth]{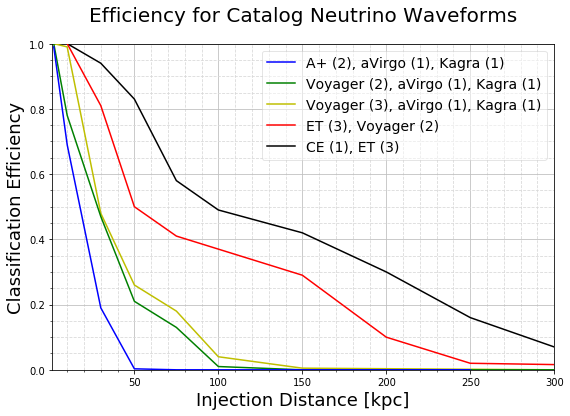}
\includegraphics[width=\columnwidth]{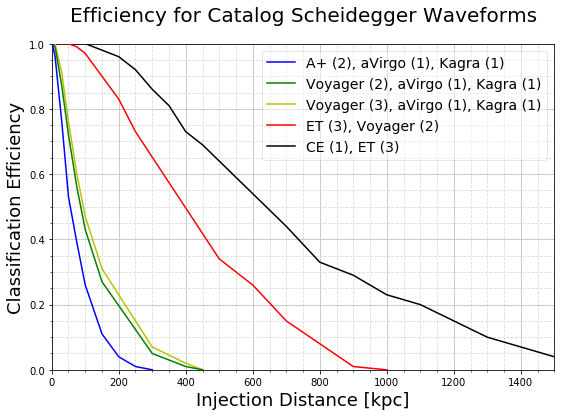}
\includegraphics[width=\columnwidth]{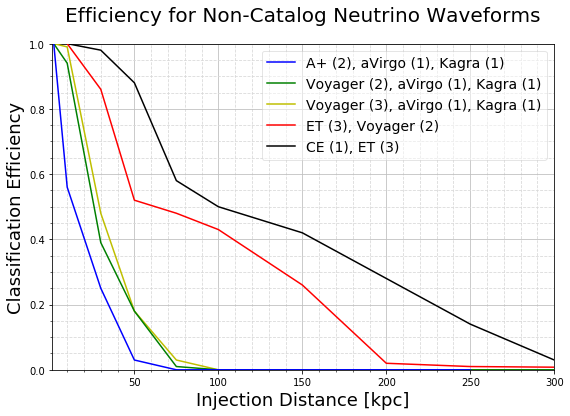}
\includegraphics[width=\columnwidth]{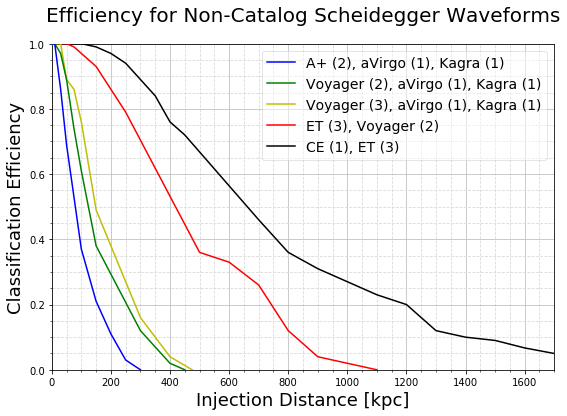}
\caption{Mechanism classification efficiency. Top plots show results for catalog waveform injections, bottom plots show results for non-catalog injections. Non-catalog injections are considered to be the most realistic test case for a genuine gravitional wave signal from an arbitrary source.}
\label{fig:eff_mech}
\end{figure*}

\begin{figure*}[p]
\centering
\includegraphics[width=\columnwidth]{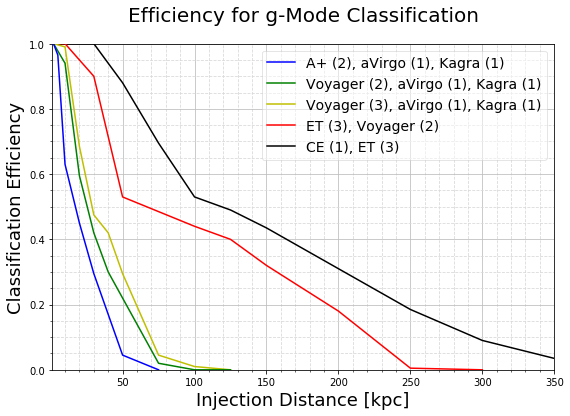}
\includegraphics[width=\columnwidth]{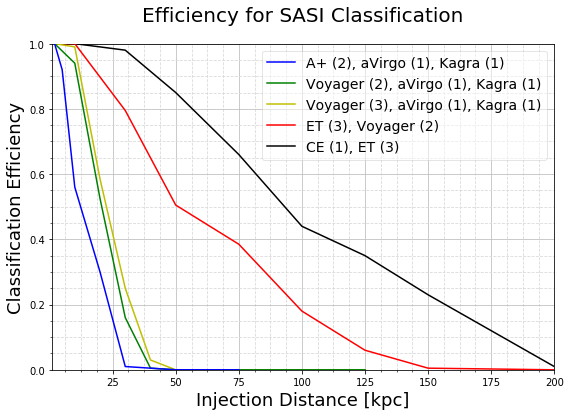}
\caption{Classification efficiency for g-mode (left) and SASI (right) waveform features. Performance was better for g-mode classification in our tests, but this is also heavily dependent on the energy of the specific waveform. Overall performance was similar to that of neutrino model mechanism classification.}
\label{fig:eff_features}
\end{figure*}

SMEE's ability to determine a source's explosion mechanism is shown in Figure~\ref{fig:eff_mech} for both catalog waveform injections and non-catalog injections. The performance differs greatly when injecting magnetorotational model waveforms vs injecting neutrino model waveforms due to the former having greater energy levels. For both explosion mechanism models, the non-catalog performance is very similar to that of catalog waveforms. 
% The small performance differences appear to be mostly due to the non-catalog waveforms and catalog waveforms having slightly different average $h_{rss}$ values. 
The fact that non-catalog performance is on par with catalog performance is a testament to the robustness of SMEE's spectrogram format. All future detector arrangements were able to confidently classify magnetorotational injections well beyond the limits of the Milky Way, with the third generation detectors confidently classifying about 10\% of injections at a distance of 1.5\,Mpc. For neutrino model waveforms, all future detector arrangements performed with greater than 56\% efficiency at 10\,kpc, suggesting that the explosion mechanism of a Galactic CCSN would likely be confidently classified. %Based off of our results, 
Low energy explosions, and explosions originating from unfortunate sky positions, could still fail to be classified from within our galaxy in LIGO's A+ configuration. Only 2-5\% of Galactic injections failed to be confidently classified in our Voyager configurations. Adding a third Voyager detector improves coverage and efficiencies at close distances, but does little to increase SMEE's range. All Galactic injections were confidently classified in the ET and Cosmic Explorer arrangements.

\subsection{Waveform Features}
\label{subsec:features_future}

SMEE's ability to detect g-modes and SASI is shown in Figure~\ref{fig:eff_features}. The figure shows the efficiencies for all injections, half of which contained the waveform feature and half of which did not. In general the performance is similar to that of neutrino model waveforms for mechanism classification. At 10\,kpc in our simulated A+ arrangement, the SASI and g-mode classification efficiencies were 63\% and 56\% respectively, suggesting that these statements would likely reach the confidence threshold for a galactic CCSN. Our results suggest that when the ET or Cosmic Explorer detectors are operational, we should be able to determine if g-modes or SASI are present in Galactic CCSN signals even if they occur in a part of the sky with poor detector sensitivity.  

%%%%%%%%%%%%%%%%%%%%%%%%%%%%%%%%%%%%%%%%%%%%%%%%%%%%%%%%%%%%%%%%%%%%%%%
%%%%%%%%%%%%%%%%%%%%%%%%%%%%%%%%%%%%%%%%%%%%%%%%%%%%%%%%%%%%%%%%%%%%%%%
\section{Conclusions}
\label{sec:conclusion}

Spectrograms offer a robust way to approach and study gravitational wave emission from a CCSN signal. The specrogram version of SMEE presented here does not use unreliable phase data. Instead, its statistics depend entirely on the power, frequency, and time of the gravitational wave emission. This is reflected in the reduction of the number of PCs needed from those in previous studies that used the time series waveforms. A reduced number of PCs results in a faster analysis, as the run time of nested sampling algorithms is proportional to the number of parameters of the signal models. 

The time-frequency path, inherent to a spectrogram, also allows the study and analysis of specific waveform features. This results in a robust and sensitive tool to perform the difficult task of parameter estimation of a gravitational wave signal from a CCSN. The small number of simulated waveforms is still a concern, as is the fact that most simulations are ended prematurely. As more simulations continue to be released they will be incorporated into SMEE's analysis.

We perform the first study of CCSN waveforms in future detector networks. SMEE's performance in future detectors was simulated and mapped out as realistically as possible with recolored aLIGO data. The results suggest that SMEE should be able to classify CCSN waveforms from well outside of the Milky Way in future detector configurations. Third generation detectors should be able to resolve the explosion mechanism and waveform features for most CCSN out to beyond the Small and Large Magellanic Clouds (50 - 60\,kpc), as well as for the dozen or so satellite galaxies in between~\cite{karachentsev2004catalog, belokurov2007cats}. A fraction of magnetorotational waveforms should be classifiable all the way out to the Andromeda galaxy (790\,kpc) in third generation detectors, while neutrino model performance falls off significantly beyond 150\,kpc. There are a total of 28 known Galaxies, mostly satellite or dwarf, within the range of 150\,kpc, and 55 within the range of 790\,kpc~\cite{karachentsev2004catalog, belokurov2007cats}. While CCSN are rare (1 - 2 per galaxy per century~\cite{1991ARA&A..29..363V, 1993A&A...273..383C, Alexeyev2002}), these results suggest that in third generation configurations a detection and accurate classification are both very plausible.

This study is the first to attempt to identify specific features associated with g-modes and SASI in detected gravitational wave signals. If more common features are found in future CCSN simulations, they could be incorporated into SMEE's analysis. The observation of a gravitational wave from a CCSN will be an important moment in astrophysics, and with a tool such as SMEE it will be possible to learn about the source and the relevant underlying physics of the CCSN explosion.

% Going forward, the parameters of SMEE's spectrograms or analysis may change. Future detectors will offer improved low frequency sensitivity, meaning that longer fast-fourier transforms might allow us to take advantage of this with more bins at low frequencies. Future iterations of SMEE may also incorporate the second polarization, cross, into its reconstructions and analysis. This would require the PCs to be complex so that both polarizations could be combined in the Fourier domain. The likelihood, and its dependence on power, would not change.

% SMEE's configuration may change as it is optimized, but it is a functioning parameter estimation tool for gravitational waveforms from CCSN, allowing us to make evidence-based statements about an observed CCSN signal. The eventual observation of a gravitational wave from a CCSN will be an important moment in astrophysics, and with a tool such as SMEE it wll be possible to learn about the source and the relevent physics of a CCSN explosion.

%%%%%%%%%%%%%%%%%%%%%%%%%%%%%%%%%%%%%%%%%%%%%%%%%%%%%%%%%%%%%%%%%%%%%%%
%%%%%%%%%%%%%%%%%%%%%%%%%%%%%%%%%%%%%%%%%%%%%%%%%%%%%%%%%%%%%%%%%%%%%%%
\begin{acknowledgments}
The authors thank Bernhard M\"uller for helpful discussions related to this work. 
JP is supported by the Australian Research Council Centre of Excellence for Gravitational Wave Discovery (OzGrav), through project number CE170100004. VR and RF are supported by National Science Foundation grant NSF PHY-1607336. ISH is supported by the Science and Technology Facilities Council grant ST/L000946/1, and the Scottish Universities Physics Alliance. The authors are grateful for computational resources provided by the LIGO Lab and supported by National Science Foundation Grants PHY-0757058 and PHY-0823459. 
\end{acknowledgments}

%%%%%%%%%%%%%%%%%%%%%%%%%%%%%%%%%%%%%%%%%%%%%%%%%%%%%%%%%%%%%%%%%%%%%%
%%%%%%%%%%%%%%%%%%%%%%%%%%%%%%%%%%%%%%%%%%%%%%%%%%%%%%%%%%%%%%%%%%%%%%

\bibliographystyle{apsrev}

\bibliography{bibfile}

\end{document}